\documentclass[10pt]{amsart}
\usepackage{amsaddr}
\usepackage{textcomp}
\usepackage{hyperref}
\usepackage{cite}
\usepackage{tikz}
\usetikzlibrary{patterns,snakes}
\usepackage{mathtools}
\usepackage{dsfont}
\usepackage{tikz}
\usepackage{graphicx}
\usepackage{dcolumn}
\usepackage{bm}

\usepackage{dcolumn}
\usepackage{subcaption}

\usepackage{bm}
\newcommand{\Brho}{\boldsymbol{\rho} }
\newcommand{\bfJ}{\boldsymbol{J} }
\newcommand{\bfz}{\boldsymbol{z} }
\newcommand{\bfrho}{\boldsymbol{\rho} }
\newcommand{\be}{\begin{equation}}
\newcommand{\ee}{\end{equation}}

\usepackage{pifont}
\expandafter\let\csname equation*\endcsname\relax
\expandafter\let\csname endequation*\endcsname\relax
\usepackage{amssymb}
\usepackage{amsmath}
\usepackage{amsthm,amsfonts}

\usepackage{epstopdf}
\usepackage{caption}
\usepackage{subcaption}
\usepackage{float}
\usepackage{enumitem}
\usepackage{calc}

\RequirePackage{graphicx}

\usepackage[margin=1in]{geometry}

\theoremstyle{plain}

\newcounter{mycounter}
\newcommand{\particles}[1]{
		\foreach \i in {0,...,11}
	{
		\draw [very thick] (\i+0.1,0) -- (\i+0.9,0);
	}
  \setcounter{mycounter}{1}
    \foreach \i in #1
    {
    \ifnum \i = 0
      \draw [very thick, dashed] (\themycounter-0.5,0.4) circle (8pt);
    \else
     \ifnum \i = 1
      \draw [very thick] (\themycounter-0.5,0.4) circle (8pt);
      \else
     \ifnum \i = 21
      \draw [very thick, red] (\themycounter-0.5,0.4) circle (8pt);
       \else
     \ifnum \i = 20
      \draw [very thick, dashed,red] (\themycounter-0.5,0.4) circle (8pt);
    \fi
    \fi
    \fi
    \fi
    \stepcounter{mycounter};
    }
}

\title{Hydrodynamics of Multi-Species Driven Diffusive Systems with Open Boundaries: A Two-Tasep Study}

\author{Ali Zahra}
\address{Laboratoire de Physique et Chimie Th\'eoriques, CNRS UMR 7019, Universit\'e de Lorraine, Boulevard des Aiguillettes -- B.P. 70239, F-54506 Vandoeuvre-les-Nancy, France}
\address{Departamento de Matem\'atica, Instituto Superior T\'ecnico, Av. Rovisco Pais, 1049-001 Lisboa, Portugal}
\email{ali.zahra@univ-lorraine.fr}

\begin{document}

\begin{abstract}
In this short note, we review a recently developed method for analysing multi-component driven diffusive systems with open boundaries. The approach generalises the extremal-current principle known for single-component models and is based on solving the Riemann problem for the corresponding hydrodynamic equations. As a case study, we focus on a two-species exclusion process on a lattice (Two-TASEP), where two types of particles move in opposite directions with two arbitrary rates and exchange positions upon encounter with a third rate. Despite its simplicity, this toy model effectively captures the key features of multi-species driven diffusive systems, including phase separation phenomena. This allows us to illustrate the critical role played by the underlying Riemann invariants in determining the system's macroscopic behavior.
\end{abstract}

\maketitle

\section{Introduction}

Understanding non-equilibrium stochastic processes is essential for unveiling the dynamics of complex systems across various scientific disciplines, including physics, chemistry, and biology \cite{gardiner1985handbook, murray2002introduction,anderson2013introduction, blythe2007stochastic}. These processes describe how systems evolve over time when they are not in thermal equilibrium, often leading to unexpected and rich behaviors even in simple models. Non-equilibrium statistical mechanics reveals phenomena such as spontaneous symmetry breaking, pattern formation, and dynamic phase transitions. \cite{spohn2012large}.
Driven diffusive systems, in particular, serve as fundamental models for exploring how external forces and interactions produce collective behavior far from equilibrium. To have a general idea, one can imagine a gas of  particles in a 1D lattice that is coupled to reservoirs from both sides. The driven aspect of the system is obtained by breaking the space symmetry through an external field so that there is a current of particles even if the two reservoirs on the boundaries are identical. Such systems are sometimes referred to as "bulk" driven in contrast with purely diffusive systems which typically have a symmetric microscopic dynamic such that the average current is null in a uniform homogeneous state. All over this article, we are concerned with the former type.

\begin{figure}[h!]
	\centering
	\includegraphics[scale=0.4]{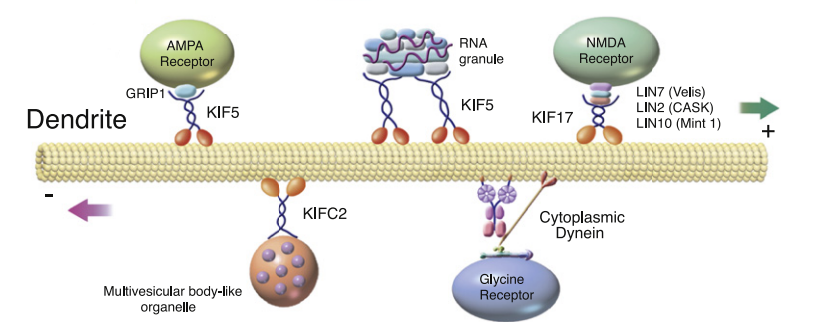} 
	\caption{Example of a multi-species driven diffusive system; molecular transport on an axon, the main nerve fiber of a neuron, featuring different types of molecular motors, each type has a preferred directed of motion and a characteristic rate. When molecules from different types meet, they interact, (by possibly slowing down), figure adapted from  \cite{hirokawa2010molecular}}
	\label{fig:multi}
\end{figure}

For systems with a single type of particles, once the expression of the current $J(\rho)$ as a function of the coarse-grained density in a homogeneous state of density $\rho$ is known, the hydrodynamic behaviour for the open-boundary system in the steady state can be determined by a simple general principle known as the extremal current principle \cite{krug1991boundary, popkov1999steady, hager2001minimal, katz_nonequilibrium_1984}. According to which, the stationary current in the steady state is given by:
\begin{equation}
\label{eqn:ecp}
j =
\begin{cases}
\max_{\rho \in [\rho^{R}, \rho^{L}]}(J(\rho))  & \text{if $\rho^{L}>\rho^{R}$} \\
\min_{\rho \in [\rho^{L}, \rho^{R}]}(J(\rho))  & \text{if $\rho^{L}<\rho^{R}$} \\
\end{cases}
\end{equation}
Where $\rho^L$, and $\rho^R$ are the densities of the left and right reservoirs respectively.
This principle doesn't only yield the current, but also allows to sketch a phase diagram. A classic example is the Totally Asymmetric Simple Exclusion Process (TASEP), where its application leads to a phase diagram featuring three distinct phases, governed by the boundary densities. For a pedagogical review, see \cite{blythe2007nonequilibrium}.
Despite its success in addressing open-boundary problems in numerous models, the extremal current principle is restricted to systems with a single type of particles and it's not clear how to generalise it to multi-species driven diffusive systems, where different types of particles interact. Such generalizations are crucial for practical applications. See Figure \ref{fig:multi} for an illustrative example.

In certain special cases of multi-species systems, exact analysis remains feasible. A notable early example is presented in \cite{evans1995asymmetric,evans1995spontaneous}, where a two-species Exclusion Process is studied under constraints on bulk and boundary rates, such that the entire system (bulk and boundaries) exhibits invariance symmetry when the two particle species and hopping directions are exchanged. This can be sometimes called, by an abuse of language, a hole-particle symmetry. The steady state becomes exactly solvable through the Matrix Product Ansatz (MPA) under additional parameter restrictions. Otherwise, a mean-field approximation is employed to analyze the system and construct a phase diagram, revealing a phase with power-law decay and another with exponential decay of the density profile. 

In another special case, it is possible to decompose the 2-species TASEP into two single-species systems by viewing the second-class particles as void for one system and first-class particles for the other system, this is sometimesmes called the coloring argument \cite{ayyer2010some}. A Colorable model is in general not integrable (in the sense that no matrix representation for the steady state exists) except for special values of the boundary rates \cite{crampe2015open}. 
	
Another example is given in \cite{khorrami2000exact} where a multi-species generalization of TASEP with open boundaries was treated exactly with MPA with hopping rates of particles drawn from a distribution with hierarchical priority. There is no hole-particles symmetry here, however, the applicability of the generalization of the MPA required that only one parameter expression is allowed for injecting particles and another for extracting particles, which results in a two parameters phase transition similar to one species TASEP for a class of distributions, while the HD phase is missing for the rest of distributions.

More generally, a quadratic algebra MPA description of the steady state of a multi-species stochastic system will always lead to a constraint on the rates of the boundaries \cite{alcaraz1998n}
	
A simplifying special situation is when the current of one of the species (or a combination of species) is null. This is the case in \cite{arita2006phase,arita2006exact} where 2-species TASEP is considered. Stationary state and phase transition are obtained,  using MPA, and are shown to be composed of three regions similar to one-species TASEP. In \cite{uchiyama2008two}  2-species ASEP with confined first class particles was analyzed and, phase transition was obtained with exact methods. This model was generalized in \cite{cantini2016koornwinder} to multi-species ASEP with again semi-permeable boundaries, phase transition was analyzed in \cite{ayyer2017exact}.
A notable recent treatment of the two-species TASEP with open boundaries is presented in \cite{cantini-open2}, while a systematic classification of the two-species ASEP with integrable boundaries is provided in \cite{crampe2015open}.

Aside from exact methods, which require fine tuning, one can turn to hydrodynamic approaches to analyze open boundary-driven diffusive systems. However, a major challenge in this context is the lack of a general method to relate boundary densities to boundary hopping rates (or other coupling constants). A notable exception occurs when the system possesses a product invariant measure for translationally invariant states. This condition not only makes mean-field currents exact but also enables a simple, explicit relation between effective boundary densities and hopping rates, given appropriate constraints on the boundary dynamics. Such systems are discussed in, for example, \cite{rakos2004exact,popkov2004spontaneous,popkov2004infinite,popkov2011hierarchy,popkov2004hydrodynamic, grosskinsky2003stationary}.

In these notes, we are reviewing the method that is introduced in \cite{cantini2024steady}. 
This method does not need integrable boundaries nor it requires a product invariant measure.  In fact, we apply it to a two-component TASEP with arbitrary inter-species hopping rates, where the invariant measure is not of product form, and the hydrodynamic currents therefore do not have a mean field form. Additionally, we choose completely arbitrary boundary hopping rates. One ingrediant of this method is the following principle

\subsection{A Principle for Multi-Component Systems}
We put forward a link between the two following problems \cite{cantini2024steady} :

\begin{itemize}
\item[$\blacktriangleright$]  \textbf{The Riemann Problem}: This involves solving a system of conservation laws, 
        $\partial_t \Brho(x,t) + \partial_x \bfJ(x,t) = 0,$
    with a step initial condition,
 $   \Brho(x,0) = \Brho^{L} \mathds{1}_{x<0} + \Brho^{R} \mathds{1}_{x>0}, \quad x \in \mathbb{R},$
    where $\Brho = (\rho_1, \ldots, \rho_N)$ represents the set of $N$-species densities, and $\bfJ(\Brho)$ denotes the associated hydrodynamic currents in a uniform state.

\item[$\blacktriangleright$]  \textbf{The Open Boundary Problem}: In this setup, the model is defined on a finite domain and coupled to reservoirs on both boundaries, with the coupling specified, for instance, through fixed boundary hopping rates for particle systems on the lattice.
\end{itemize}

For the second problem, the boundary densities are often unknown. However, if these densities can be determined (e.g., through measurement), the principle in \cite{cantini2024steady} states that the bulk densities correspond to the solution of the Riemann problem at 
$x=0$, using a step initial condition where the left and right densities are given by the two reservoir densities, respectively.

For single-species models, this principle is essentially a reformulation of the extremal current principle, as shown in \cite{cantini2024steady}. Let's give the elementary example of single species TASEP that will serve as well to introduce new terminology which are compared to the standard terminologies in Fig. \ref{TASEPopen} and  will be relevant for the multispecies case.

\subsection{Elementary Example: Single-Species TASEP}
Both the Riemann and open boundary problems are well known for TASEP (see \cite{blythe2007nonequilibrium} for a review). The hydrodynamic current is given by \(J(\rho) = \rho(1-\rho)\), which gives rise to the non-viscous Burgers equation as the conservation law, $\partial_t \rho(x,t) + (1-2 \rho) \partial_x \rho(x,t) = 0$ The solution to the Riemann problem is a function of $v := \frac{x}{t}$ and reads:
\begin{itemize}
    \item If \(\rho^L > \rho^R\), a rarefaction fan solution:  
$    \rho(v) = 
    \begin{cases} 
    \rho^L, & v < 1 - 2\rho^L, \\
    \frac{1-v}{2}, & 1 - 2\rho^L < v < 1 - 2\rho^R, \\
    \rho^R, & v > 1 - 2\rho^R,
    \end{cases}$
    \item If \(\rho^L < \rho^R\), a shock solution:  
$    \rho(v) = 
    \begin{cases} 
    \rho^L, & v < 1 - \rho^L - \rho^R, \\
    \rho^R, & v > 1 - \rho^L - \rho^R,
    \end{cases}$
\end{itemize}

For TASEP with open boundaries, note that the boundary densities are trivially related to the boundary rates: \(\rho^L\) is equal to the particle injection rate on the left, and \(1 - \rho^R\) corresponds to the particle extraction rate on the right. An important quantity for the analysis is the characteristic velocity in the bulk, defined as:
$v^B := J'(\rho^B) = 1 - 2\rho^B.$
Each phase of the model is distinguished either a sign of \(v^B\), or by \(v^B\) being null. To facilitate future analysis, we introduce new phase terminologies:

\begin{itemize}
    \item \textbf{Left-Induced Phase (LI)}: The bulk density equals the left boundary density, typically referred to as the low-density phase. In this phase, \(v^B > 0\), indicating that perturbations travel from the left boundary toward the bulk.

    \item \textbf{Right-Induced Phase (RI)}: The bulk density equals the right boundary density, typically referred to as the high-density phase. In this phase, \(v^B < 0\), indicating that perturbations travel from the right boundary toward the bulk.

    \item \textbf{Bulk-Induced Phase (BI)}: The bulk density is neither equal to the left nor the right boundary densities. For this phase, \(v^B = 0\), and the bulk density thus the solution of the equation $v^B(\rho^B)=0$. i.e. \(\rho^B = \frac{1}{2}\) in this model.
\end{itemize}

	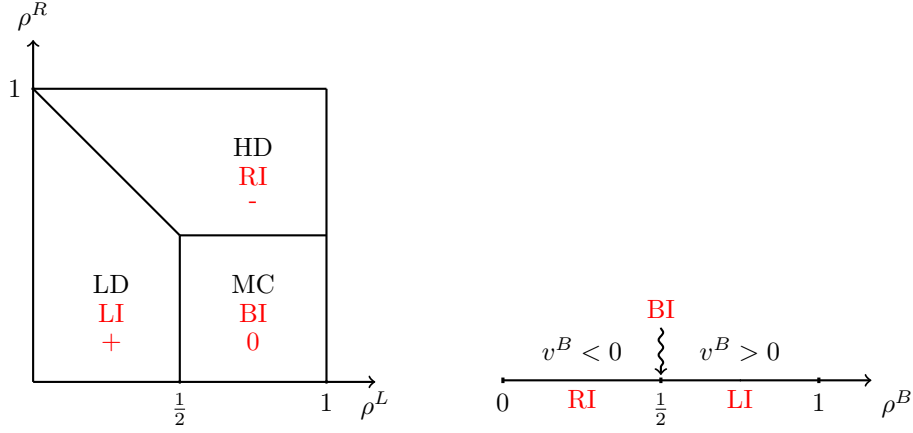
\begin{figure}[h!]
		\centering
		\begin{subfigure}[b]{0.4\linewidth}
			\begin{center}
				\begin{tikzpicture}[thick, scale=0.65]
				\draw[thick,->] (0,0) -- (7,0) node[anchor=north] {$\rho^{L}$};
				\draw[thick,->] (0,0) -- (0,7) node[anchor=south ]{$\rho^{R}$};	
				
				\draw (3 cm,1pt) -- (3 cm,-1pt) node[anchor=north] {$\frac{1}{2}$};
				
				\draw (6 cm,1pt) -- (6 cm,-1pt) node[anchor=north] {$1$};

				\draw (1pt,6 cm) -- (-1pt,6 cm) node[anchor=east] {$1$};
				
				\draw (3,3) -- (6,3);
				
				\draw (0,6) -- (6,6);
				\draw (0,6) -- (3,3);
				\draw (3,0) -- (3,3);
				\draw (6,0) -- (6,6);
				
				\draw (5.7,1) node[anchor=east] {};
				
				\draw (4.5,5.2) node[anchor=north] {HD};
				\draw (4.5,4.6) node[anchor=north] {\textcolor{red}{RI}};
                \draw (4.5,4) node[anchor=north] {\textcolor{red}{-}};
				
				\draw (1.6,2.4) node[anchor=north] {LD};
				\draw (1.6,1.8) node[anchor=north] {\textcolor{red}{LI}};
                \draw (1.6,1.2) node[anchor=north] {\textcolor{red}{+}};

				\draw (4.5,2.4) node[anchor=north] {MC};
				\draw (4.5,1.8) node[anchor=north] {\textcolor{red}{BI}};
                \draw (4.5,1.2) node[anchor=north] {\textcolor{red}{0}};
				
				\end{tikzpicture}
			\end{center}
		\end{subfigure}
		\begin{subfigure}[b]{0.4\linewidth}
			\begin{center}
				\begin{tikzpicture}[thick, scale=0.7]
				\draw[thick,->] (0,0) -- (7,0) node[anchor=north west] {$\rho^{B}$};

                \draw[thick ,-> ,decorate,decoration={snake,amplitude=.3mm,segment length=2mm,post length=1mm}] (3 cm,1 cm) node[anchor=south] {\textcolor{red}{BI}} -- (3 cm, 0.1 cm) ;
				
				\draw [very thick] (3 cm,1.5pt) -- (3 cm,-1.5pt) node[anchor=north] {$\frac{1}{2}$};
				
				\draw [very thick] (6 cm,1.5pt) -- (6 cm,-1.5pt) node[anchor=north] {$1$};
				\draw [very thick] (0 cm,1.5pt) -- (0 cm,-1.5pt)  node[anchor=north] {$0$};
				
				\draw (1.5 cm,0pt) node[anchor=north] {\textcolor{red}{RI}};
\draw (1.5 cm, 1 cm) node[anchor=north] {$v^B < 0 $};

				\draw (4.5 cm,0pt) -- (4.5 cm,0pt)
                node[anchor=north] {\textcolor{red}{LI}};

           	\draw (4.5 cm,1 cm) 
                node[anchor=north] {$v^B >0 $};
				
				\end{tikzpicture}
			\end{center}
		\end{subfigure}
	\caption{Phase diagram of a single-species TASEP, shown in terms of boundary densities on the left and bulk density on the right. Acronyms in black represent the conventional phase names: HD (High Density), LD (Low Density), and MC (Maximal Current). Acronyms in red indicate the new phase names: RI (Right Induced), LI (Left Induced), and BI (Bulk Induced). The symbols ${+, 0, -}$ refer to the characteristic velocities in the bulk.}
		\label{TASEPopen}
	\end{figure}

As depicted in Figure~\ref{TASEPopen}, the phase diagram can be sketched using the bulk density as an order parameter, with the phases distinguished by the sign of \(v^B\). One can also verify that, for each phase, the bulk density matches the solution at \(x=0\) for the corresponding Riemann problem.

The Riemann problem is generally easier to solve than the open-boundary problem. Section \ref{Hyperbolic} provides an operational overview of the key tools from the theory of conservation laws, so readers unfamiliar with these ideas are encouraged to start there. In particular, one should be comfortable with the notions of shock and rarefaction solutions, shock and rarefaction curves, the Rankine–Hugoniot condition, admissibility criteria, and the use of Riemann variables.

However, solving the Riemann problem alone is not sufficient to solve the general multi-species open boundary problem, as the boundary densities are typically unknown. In Section \ref{Open-boundary}, we illustrate how to overcome this limitation and determine both the bulk and boundary densities using the 2-TASEP as an example. Based on this analysis, we also discuss the construction of the phase diagram for the 2-TASEP in the same section. Section \ref{hydro} provides a detailed review of the solution to the Riemann problem for the 2-TASEP, summarizing key results from \cite{cantini2022hydrodynamic}.

\section{Hydrodynamics for the Two-TASEP}
\label{hydro}
Defined on the lattice and  thought of as two TASEPs propagating in opposite directions, one denoted $\bullet$ jumping only to the right at a rate $\beta$ and the other, denoted $\circ$ jumping only to the left at a rate $\alpha$. when two particles of different types meet, they swap at rate that can be fixed to $1$ by a change of time scale. If we denote the void particles as $\ast$, the local microscopic rules are summarized:

\begin{equation}
    \begin{split}
	\bullet\,\ast& ~ \rightarrow ~ \ast\,\bullet\quad\text{rate}\quad \beta\\
	\ast\,\circ& ~ \rightarrow ~ \circ\,\ast\quad\text{rate}\quad \alpha\\
	\bullet\,\circ& ~ \rightarrow ~ \circ\,\bullet\quad\text{rate}\quad 1
\end{split}
\end{equation}

\begin{figure}
    \centering
    \includegraphics[width=0.9\linewidth]{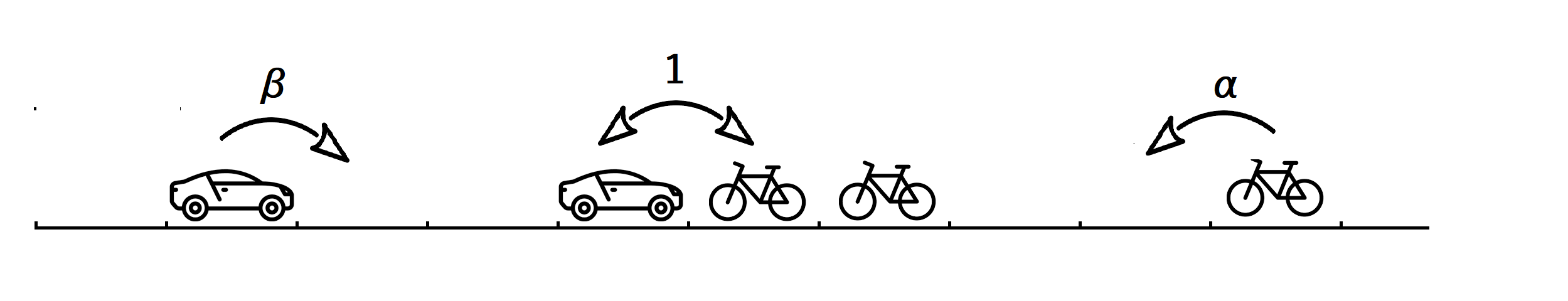}
    \caption{The two-TASEP as a traffic flow model, $\bullet$ particles represent cars, $\circ$ particles represent bikes}
    \label{trafic}
\end{figure}
This model is closely related to several others in the literature~\cite{popkov2002sufficient,arndt1998spontaneous,cantini2008algebraic}.
Figure~\ref{trafic} provides a traffic-flow interpretation.  
In the special case $\alpha = \beta = 1$, the $\ast$ particles can be identified as second-class particles, the $\bullet$'s as first-class, and the holes as $\circ$ particles~\cite{ferrari1994invariant}.

Unlike the single-species TASEP, the invariant measure of the two-species TASEP is not generally of product form.  
An exception occurs in the case $\alpha = \beta = \frac{1}{2}$, where a product-form invariant measure was shown to exist~\cite{fritz2004derivation}.  
In the same work, the hydrodynamic limit was rigorously derived using Yau’s relative entropy method.  
This result extends to the broader class of models satisfying $\alpha + \beta = 1$, for which the hydrodynamic equations can be mapped—via a similarity transformation—back to the Leroux system~\cite{cantini2022hydrodynamic,fritz2004derivation}.  
See also the related discussion in~\cite{toth2003onsager}.

Apart from this special case, the hydrodynamic currents do not have a mean-field form. These currents were computed in the hydrodynamic limit of a system with periodic boundary conditions using Nested Algebraic Bethe Ansatz, \cite{cantini2008algebraic}, and they read:
	\begin{gather}
 \label{1Jrz}
	J_\circ= z_\alpha(z_\beta-1)+\rho_\circ(z_\alpha-z_\beta)\\
	\label{2Jrz}
	J_\bullet= z_\beta(1-z_\alpha)+\rho_\bullet(z_\alpha-z_\beta)
	\end{gather}
 
where the parameters $z_\alpha$, $z_\beta$ belong to a physical domain $D_z$ such that $z_\alpha \in [0,\min(1,\alpha)]$, $z_\beta \in [0,\min(1,\beta)]$ and $z_\alpha + z_\beta \leq 1$. These parameters are functions of the densities, given implicitly as solution of the equations:
	\begin{gather}
 \label{rho-z0}
	\frac{\rho_\circ}{z_\alpha}+\frac{\rho_\bullet}{z_\alpha-1}+\frac{1-\rho_\circ-\rho_\bullet}{z_\alpha-\alpha}=0\\ \label{rho-z1}
	\frac{\rho_\bullet}{z_\beta}+\frac{\rho_\circ}{z_\beta-1}+\frac{1-\rho_\circ-\rho_\bullet}{z_\beta-\beta}=0.
	\end{gather}

In the appendix \ref{Annex}, we write the explicit mapping between $(z_\alpha,z_\beta)$ and $(\rho_{\bullet},\rho_{\circ})$, as well as the explicit expression of the currents. However, we will see that the implicit expression is indeed conceptually advantageous: fixing $z_{\alpha}$ in \ref{rho-z0} defines a linear relation of densities, that we represent by the line $l_{\alpha}$ in the density domain.
Similarly, through eq.\ref{rho-z1}. define the line $\l_{\beta}$ of densities sharing the same $z_{\beta}$.
We will soon see that: (a)- these lines turn out to simply be the shock curves.
(b)- The Riemann variables turn out to be the $\bfz$ variables. First notice that we can write the currents as solutions of a system of two linear equations by
solving eqs.\ref{1Jrz},\ref{2Jrz} for the densities and plugging them into eqs.\ref{rho-z0},\ref{rho-z1}
\begin{gather}\label{Jz0}
\frac{J_\circ}{z_\alpha}+\frac{J_\bullet}{z_\alpha-1}-\frac{J_\circ+J_\bullet}{z_\alpha-\alpha}+1=0\\\label{Jz1}
\frac{J_\bullet}{z_\beta}+\frac{J_\circ}{z_\beta-1}-\frac{J_\circ+J_\bullet}{z_\beta-\beta}+1=0.
\end{gather}
\subsubsection*{Shock solutions:}
To investigate the structure of the shocks, (see section \ref{cons}) assume $x_s$ to be the position  shock of a shock, and denote $[{\bfrho}]$ and $[{\bf J}]$ the discontinuity jumps of the densities and of the currents respectively across the shock:
$$
[{\bfrho}]=\bfrho(x_s^+)-\bfrho(x_s^-),\qquad [{\bf J}]={\bf J}(x_s^+)-{\bf J}(x_s^-)
$$

Eqs. \ref{rho-z0},\ref{Jz0} translate into:

\begin{equation}
\begin{split}
  \alpha (1- z_\alpha) [\rho_\circ] + (1 - \alpha) 
 z_\alpha [\rho_\bullet] = 0 \\
\alpha (1 -z_\alpha)  [J_\circ]  + (1 - \alpha)z_\alpha [J_\bullet]  = 0
\end{split}
\end{equation}

This means that whenever $\Brho^{+}$ and $\Brho^{-}$ belong to a line $l_{\alpha}$, we have the Rankin-Hugoniot condition verified $\frac{[J_\bullet]}{[\rho_\bullet]}
=
\frac{[J_\circ]}{[\rho_\circ]}$, implying a shock solution between $\Brho^{+}$ and $\Brho^{-}$. In the same manner, we can combine eq. \ref{rho-z1} and eq. \ref{Jz1} to establish that $\l_{\beta}$ is a shock curve. What is left is to apply a stability condition to select the physical shocks, this amount to adding a direction to the shock curves. In conclusion, we have the two types of admissible shocks
\begin{itemize}
	\item $\beta$-shocks, featuring a discontinuity only of $z_{\beta}$; $z_\alpha^+=z_\alpha^-$ with shock velocity function of the form: $v_{s,\beta}(z_\alpha;z_\beta^-,z_\beta^+)$, and an admissibility condition: $z_{\beta}^{-} < z_{\beta}^{+}$
	\item $\alpha$-shocks: the other way around. The velocity function is of the form $v_{s,\alpha}(z_\beta;z_\alpha^-,z_\alpha^+)$, and the admissibility condition:  $z_{\alpha}^{-} > z_{\alpha}^{+}$
\end{itemize}
The explicit expression of the functions $v_{s,\alpha}$,$v_{s,\beta}$ are cumbersome. We provide additional insights into their computations in the appendix \ref{Annex}.

Since the shock curves are straight lines, that leads automatically to conclude that they are as well the rarefaction curves, and that the model belongs to the Temple class.

These properties allow as well to easily identify the $\bfz$ variables as the Riemann invariants.
We know that the rarefaction curve $l_{\alpha}$ is 
(a)- tangent to the right eigenvector $r_{\alpha}$ which is orthogonal to the left eigenvector $ll_{\beta}$
(b)- since it's a level curve for the scalar field $z_{\alpha}$, it is orthogonal to the gradient of $z_{\alpha}$.
From the two previous remarks, we conclude that the left eigenvector associated with the field $z_\alpha$ is indeed co-linear to the gradient of $z_\alpha$, hence, $z_\alpha$ can be seen as a Riemann invariant.

We can show in another way that the $\bfz$ variables are the Riemann variables by deriving "partially" decoupled conservation equations in terms of these variables: differentiate the l.h.s. of eq.(\ref{rho-z0}) with respect to $t$, differentiate the l.h.s. of eq.(\ref{Jz0}) with respect to $x$ and sum the obtained results.

Thanks to the conservation laws, the derivatives $\partial_t \bfrho$ and $\partial_x \bfJ$ cancel and 
one remains with
\begin{equation}\label{conserv-za}
\partial_t z_\alpha +v_\alpha(\bfz) \partial_x z_\alpha=0,\qquad v_\alpha(\bfz)=\frac{\left(\frac{J_\circ}{z_\alpha^2}+\frac{J_\bullet}{(z_\alpha-1)^2}-\frac{J_\circ+J_\bullet}{(z_\alpha-\alpha)^2}\right)}{\left(\frac{\rho_\circ}{z_\alpha^2}+\frac{\rho_\bullet}{(z_\alpha-1)^2}+\frac{1-\rho_\circ-\rho_\bullet}{(z_\alpha-\alpha)^2}\right)}.
\end{equation}
In the same way one obtains the equation for $z_\beta$
\begin{equation}\label{conserv-zb}
\partial_t z_\beta +v_\beta(\bfz) \partial_x z_\beta=0,\qquad v_\beta(\bfz)=\frac{\left(\frac{J_\bullet}{z_\beta^2}+\frac{J_\circ}{(z_\beta-1)^2}-\frac{J_\circ+J_\bullet}{(z_\beta-\beta)^2}\right)}{\left(\frac{\rho_\bullet}{z_\beta^2}+\frac{\rho_\circ}{(z_\beta-1)^2}+\frac{1-\rho_\circ-\rho_\bullet}{(z_\beta-\beta)^2}\right)}.
\end{equation}
The speeds $v_\alpha$ and $v_\beta$ are the eigenvalues of the linearization matrix $\partial_{\rho_j}J_i$, and on general grounds they can also be written as
\begin{equation}
v_\alpha= \partial_{\rho_i}J_i|_{z_\beta}=\frac{\partial_{z_\alpha}J_i(\bfz)}{\partial_{z_\alpha}\rho_i(\bfz)},\qquad v_\beta= \partial_{\rho_i}J_i|_{z_\alpha}=\frac{\partial_{z_\beta}J_i(\bfz)}{\partial_{z_\beta}\rho_i(\bfz)}
\end{equation}

For explicit formulas of the characteristic velocities, see appendix \ref{Annex}.

\subsubsection*{Continuous solutions and the Rieman problem:}

Consider the Riemann problem consisting of eqs. \ref{conserv-za},\ref{conserv-zb} with the initial condition:
$$\boldsymbol{z}(x,0) = \boldsymbol{z}^{L} \mathds{1}_{x<0}(x) + \boldsymbol{z}^{R}\mathds{1}_{x>0}(x), \; x \in \mathbb{R}$$ The problem is invariant under the rescaling $(x,t) \rightarrow (\lambda x, \lambda t)$ for any $\lambda>0$, so we can write eqs. \ref{conserv-za},\ref{conserv-zb} in terms of one variable $u=\frac{x}{t}$:

\begin{equation}
    \frac{d z_{\alpha}}{d u} [-u+v_{\alpha}(z_{\alpha},z_{\beta})]=0
\end{equation}
\begin{equation}
    \frac{d z_{\beta}}{d u} [-u+v_{\beta}(z_{\alpha},z_{\beta})]=0
\end{equation}

\begin{figure}[htbp]
    \centering
    \begin{subfigure}[b]{0.45\textwidth}
        \centering

\begin{tikzpicture}[scale= 0.6, shift={(8cm,2cm)}]
		\begin{scope}[scale = .8, xshift=10cm]
			\fill [fill=white!85!blue] (-1,-1)--(-1,7) node [left] {\scriptsize $1$}--(7,-1)node [below] {\scriptsize $1$}--cycle;
			\draw[->] (-1,-1) -- (8,-1) node[below] {\scriptsize $z_{\alpha}$};
			\draw[->] (-1,-1) -- (-1,8) node[left] {\scriptsize  $z_{\beta}$};
			\begin{scope}[shift={(-1,.5)}]
				\draw [<-,>=latex,blue](2.5,3) --(1,3);
				\fill (1,3) circle (.08) node[left] {\scriptsize \bf L};
				\fill (2.5,3) circle (.08) node[right] {\scriptsize \bf R};
				
				\draw [->,>=latex,thick,black!25!green](4.1,2.4) --(5.5,1);
                \fill (4.1,2.4) circle (.08) node[left] {\scriptsize \bf L};
				\fill (5.5,1) circle (.08) node[right] {\scriptsize \bf R};
    
				\draw[<-,>=latex,red](4.5,-0.5)--(4.5,1);
				\fill (4.5,1) circle (.08) node[above] {\scriptsize \bf L};
				\fill (4.5,-0.5) circle (.08) node[below] {\scriptsize \bf R};
				\draw [->,>=latex,dashed,blue](2.5,2) --(1,2);
				\fill (1,2) circle (.08) node[left] {\scriptsize \bf R};
				\fill (2.5,2) circle (.08) node[right] {\scriptsize \bf L};
				\draw[->,>=latex,dashed,red](3.5,-0.5)--(3.5,1);
				\fill (3.5,1) circle (.08) node[above] {\scriptsize \bf R};
				\fill (3.5,-0.5) circle (.08) node[below] {\scriptsize \bf L};
			\end{scope}
		\end{scope}
	\end{tikzpicture}

        \caption{Elementary solutions}
        \label{fig:elementary}
    \end{subfigure}
    \hfill
    \begin{subfigure}[b]{0.45\textwidth}
        \centering
\begin{tikzpicture}[scale= 0.8]
	\begin{scope}[scale = .6, xshift=10cm]
		\fill [fill=white!85!blue] (-1,-1)--(-1,7) node [left] {\scriptsize $1$}--(7,-1)node [below] {\scriptsize $1$}--cycle;
		\draw[->] (-1,-1) -- (8,-1) node[below] {\scriptsize $z_{\alpha}$};
		\draw[->] (-1,-1) -- (-1,8) node[left] {\scriptsize  $z_{\beta}$};
		\begin{scope}[shift={(-1,.5)}]
			\draw [->,>=latex,dashed,blue](2,2) --(1,2);
			\draw [->,>=latex,dashed,red](1,2)--(1,3);
			\draw [->,>=latex,thick,red](1,2)--(1,1);
			\draw [->,>=latex,thick,blue](2,2) --(3,2);
			\draw [->,>=latex,thick,blue](2,2) --(4.5,2);
			\draw [->,>=latex,thick,black!25!green](4.5,2) --(5.5,1);
			\draw [->,>=latex,thick,red](5.5,1)--(5.5,0);
			\draw [->,>=latex,thick,red](3,2)--(3,1);
			\draw [->,>=latex,dashed,red](3,2)--(3,3);
			\fill (2,2) circle (.08) node[above] {\scriptsize \bf L};
			\fill (1,3) circle (.08) node[above] {\scriptsize \bf R};
			\fill (1,1) circle (.08) node[below] {\scriptsize \bf R};
			\fill (3,1) circle (.08) node[below] {\scriptsize \bf R};
			\fill (5.5,0) circle (.08) node[below] {\scriptsize \bf R};
			\fill (3,3) circle (.08) node[above] {\scriptsize \bf R};
		\end{scope}
		
	\end{scope}
\end{tikzpicture}

        \caption{Global solutions}
        \label{fig:global}
    \end{subfigure}
    \caption{Representation of the solutions to the Riemann problem in the \( \bfz \) domain. Points \( L \) and \( R \) correspond to \( (z^L_\alpha, z^L_\beta) \) and \( (z^R_\alpha, z^R_\beta) \), respectively. The continuous line represents the trajectory of a rarefaction fan, while the dashed line represents a shock solution. The left panel illustrates elementary solutions that respect the respective constraints on the initial conditions, whereas the right panel shows global solution trajectories constructed by stitching together segments of elementary solutions.}
    \label{fig:sol}
\end{figure}
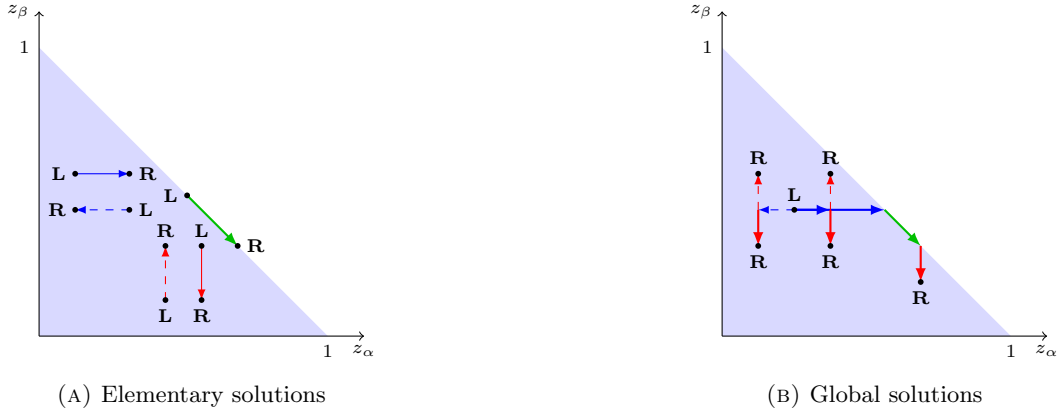

A part from trivial constant solutions for both $z_{\alpha}$ and $z_{\beta}$, we can have three types of continuous elementary solutions:

\begin{itemize}
    \item $z_{\alpha}$ is constant and $z_{\beta}(u)$ is a solution of the equation
\begin{equation}
    u = v_{\beta}(z_{\alpha},z_{\beta}(u))
\end{equation}
    We call such solution \textit{$\beta$-rarefaction fan}. This requires the initial condition to satisfy (a) $z_{\alpha}^L = z_{\alpha}^R$ and (b) $z_{\beta}^L > z_{\beta}^R$. The condition (b) is a consequence of $v_{\beta}(z_{\alpha},z_{\beta})$ being a decreasing function of $z_{\beta}$.
    We obviously need $u^L < u^R$, with
    $u^{L\backslash R} := v_{\beta}(z_{\alpha}^{L\backslash R},z_{\beta}^{L\backslash R})$
    such that the rarefaction fan is defined on $[u^L,u^R]$. Outside this domain, the solution is simply constant. This elementary solution is represented in figure \ref{fig:elementary} by a continuous vertical arrow directed downwards. Note that in this particular case of a Temple class model \cite{temple1983systems}, the condition (b) means simply that the corresponding shock solution is unstable as it violates the entropy condition. This situation is analoguous to the scalar case but not generally true for models which don't fall in the Temple class.
    \item Similarly, we can have an \textit{$\alpha$-rarefaction fan} as an elementary solution, when $z_{\beta}^L=z_{\beta}^R$ and $z_{\alpha}^L < z_{\alpha}^R$. Such a solution is given implicitly by:
    \begin{equation}
        u=v_{\beta}(z_{\alpha}(u),z_{\beta})
    \end{equation}
    This solution is represented in figure \ref{fig:elementary} as a continuous horizontal arrow directed to the right.
    
\item \( v_\alpha(z_{\alpha}, z_{\beta}) = v_\beta(z_{\alpha}, z_{\beta}) = u \). This is possible only when \( z_\alpha + z_\beta = 1\), i.e., along the diagonal boundary of the \( \bfz \) domain. Direct computation gives: 
\( v_{\alpha} = v_{\beta} = -1 + 2z_{\alpha} = 1 - 2z_{\beta} \). 
This gives rise to a third-type rarefaction fan that exists on the diagonal and is then given explicitly by:
\begin{equation}
z_{\alpha} = \frac{u}{2} + \frac{1}{2}, \quad z_{\beta} = -\frac{u}{2} + \frac{1}{2}.
\end{equation}
This solution is represented by the green arrow in Figure~\ref{fig:elementary}. Its existence as an elementary solution requires that the initial condition satisfies:
$
z^{L\backslash R}_\alpha + z_\beta^{R\backslash L} = 1, \quad z_\alpha^L < z_\alpha^R.
$
\end{itemize}

The fact that the equation \( v_\alpha(\bfz) = v_\beta(\bfz) \) has solutions implies that our system of conservation laws is not strictly hyperbolic and exhibits a continuous degeneracy along the diagonal boundary of the \( \bfz \) domain. The mapping between the densities and the Riemann variables is a simple identity on that diagonal. Therefore, this rarefaction fan corresponds to the single-species TASEP. This is not surprising, as on the diagonal of the density domain \( \rho_{\circ} + \rho_{\bullet} = 1 \), the density \( \rho_{*} = 0 \), and the model reduces to a single-species system.

\subsubsection*{Global solutions:} Each elementary solution (shocks and rarefaction fans) is valid under specific constraints on the initial conditions. For arbitrary $\bfz^L$ and $\bfz^R$, the Riemann problem can be solved by "stitching" together segments of elementary solutions in a way that satisfies the boundary requirements of each.

Diagrammatically (see Figure~\ref{fig:global}), starting from $z^L$, any point $z^R$ in the $\bfz$ domain can be reached using a unique combination of the arrows. There are five scenarios for the global solutions:
\begin{itemize}[label=\raisebox{0.2ex}{\tiny$\blacktriangleright$}]
    \item A shock of type $\alpha$ followed by a rarefaction fan of type $\beta$ if $\bfz^L$ is located north-east of $\bfz^R$. 
    \item A shock of type $\alpha$ followed by a shock of type $\beta$ if $\bfz^L$ is located south-east of $\bfz^R$. 
    \item A rarefaction fan of type $\alpha$ followed by a rarefaction fan of type $\beta$ if $\bfz^L$ is located north-west of $\bfz^R$. 
    \item A rarefaction fan of type $\alpha$, followed by a TASEP-like rarefaction fan, followed by a rarefaction fan of type $\beta$ if $\bfz^L$ is located north-west of $\bfz^R$ and $z_\alpha^R + z_\beta^L > 1$. 
    \item A rarefaction fan of type $\alpha$ followed by a shock of type $\beta$ if $\bfz^L$ is located south-west of $\bfz^R$ and $z_\alpha^R + z_\beta^L < 1$. 
\end{itemize}

\section{The open boundary problem}
\label{Open-boundary}
We consider in this section the two-TASEP model defined on a finite lattice with $L$ sites and open boundary conditions. Particles can be injected or extracted at the first and last sites, with rates that depend on the particle type:

	\begin{center}
		
		\setlength{\tabcolsep}{10pt} 
		\renewcommand{\arraystretch}{1.5} 
		\begin{tabular}{l|c|r|c} 
			\textbf{Hopping} & \textbf{Left} & \textbf{Bulk} & \textbf{Right}\\
			\hline
			\hline
			$ \bullet\,\ast \rightarrow  \ast\,\bullet $ & $ \nu_{\bullet \ast}^{L} $ & $ \beta $ &  $ \nu_{\bullet \ast}^{R} $\\
			$ \ast\,\circ \rightarrow  \circ\,\ast $ & $ \nu_{\ast \circ}^{L} $ & $ \alpha $ &  $ \nu_{\ast \circ}^{R} $\\
			$ \bullet\,\circ \rightarrow  \circ \,\bullet $ & $ \nu_{\bullet \circ}^{L} $ & $ 1 $ &  $ \nu_{\bullet \circ}^{R} $\\
		\end{tabular}
		
	\end{center}

Due to the boundary conditions and additional parameters, analyzing this setup is more challenging than the model defined on the infinite lattice $\mathbb{Z}$. A key distinction is that this system reaches a steady-state distribution over time, commonly referred to as a Non-Equilibrium Steady State (NESS).

Some fundamental questions that can be posed about this system include:

\begin{enumerate} [label=\alph*)] \item What is the current of each species in the steady state? \item What are the bulk and boundary densities? \item Are there any phase transitions induced by the boundaries, and how can they be described? \item How can fluctuations around the steady state be evaluated? \item Is the steady state exactly solvable? \end{enumerate}

The method we are reviewing focuses on addressing the first three questions. Its applicability requires the following ingredients:

\begin{itemize}
    \item The hydrodynamic currents: there are the currents of the model in uniform density stationary state. For driven diffusive systems describing Asymmetric models, they are functions of the densities. For the two TASEP, they are given by eqs.\ref{1Jrz}, \ref{2Jrz}

    \item The solution of Riemann problem for an arbitrary step initial condition $\Brho^L - \Brho^R$. In particular we need the solution at the origin. This value is independant of time and only a function of $\Brho^L$ and $\Brho^R$. Denote this function $R_0(\Brho^L,\Brho^R)$.
    This is done in the previous section for the 2-TASEP.
    
    \item The relation between the boundary densities and the boundary currents. For the two TASEPs, these relations are straightforward:
    	\begin{equation}
	\begin{split}
	& J_{\bullet}^{L} = \nu_{\bullet \circ}^{L} \rho_{\circ}^{L} + \nu_{\bullet \ast}^{L} (1- \rho_{\circ}^{L} - \rho_{\bullet}^{L} ) \\
	& J_{\circ}^{L} = -(\nu_{\bullet \circ}^{L} + \nu_{\ast \circ}^{L}) \rho_{\circ}^{L} \\
	& J_{\circ}^{R} = -\nu_{\bullet \circ}^{R} \rho_{\bullet}^{R} - \nu_{\ast \circ}^{R} (1- \rho_{\circ}^{R} - \rho_{\bullet}^{R} ) \\
	& J_{\bullet}^{R} = (\nu_{\bullet \circ}^{R} + \nu_{\bullet \ast}^{R}) \rho_{\bullet }^{R} \\
	\end{split}
	\end{equation}
Where we define the boundary densities simply as the averaged densities of the first and last site of the lattice. (conceptually, one can view these extreme sites as part of the reservoir).
\end{itemize}
In the steady state, the current of each species is constant across the lattice. For the 2-TASEP we get:

\begin{align}
\label{steady}
    J_\bullet^L(\rho_\circ^L, \rho_\bullet^L) &= J_\bullet(\rho_\circ^B, \rho_\bullet^B) = J_\bullet^R(\rho_\circ^R, \rho_\bullet^R), \notag \\
    J_\circ^L(\rho_\circ^L, \rho_\bullet^L) &= J_\circ(\rho_\circ^B, \rho_\bullet^B) = J_\circ^R(\rho_\circ^R, \rho_\bullet^R).
\end{align}

We emphasise that the boundary desnities are in general unknowns, on equal footing as the bulk densities; In practice, what is fixed is only the boundary hopping rates. No knowledge of the relationship between the boundary rates and densities is assumed.
However, there is a relationship between the boundary and the bulk densities, as explained in the introduction:

\begin{equation}
    \Brho^B = R_0(\Brho^L,\Brho^R)
\end{equation}

For 2-TASEP, this constitutes two equations relating the bulk and boundary densities. Together with \ref{steady} constitutes six equations with six variables that can be solved numerically using a simple iterative scheme; see \cite{cantini2024steady} for additional details.

\subsection{Phase Diagram for the Two-TASEP}

\begin{figure}[h!]
\label{phase}
	\centering
\includegraphics[width=0.3\linewidth]{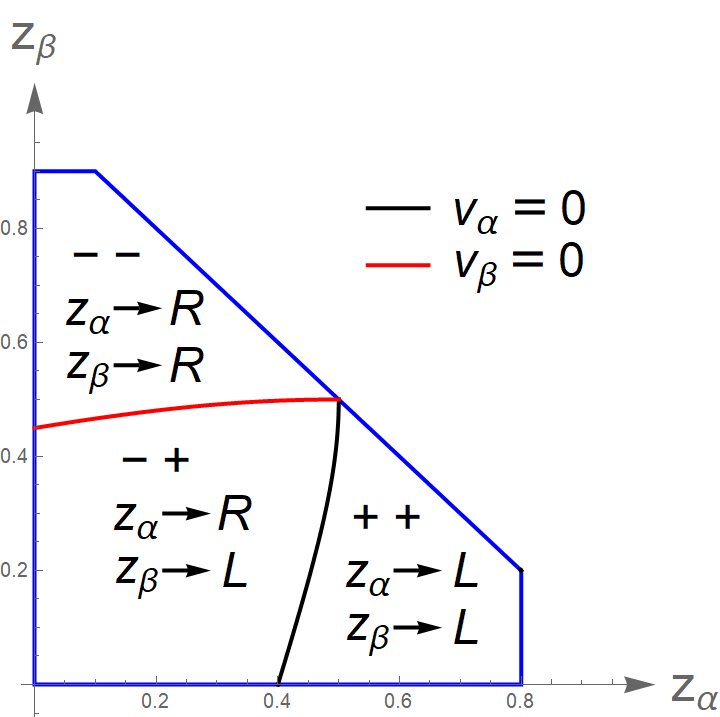}
	\caption{Phase diagram of the Two-TASEP: The signs correspond to the velocities $v_\alpha$ and $v_\beta$ in order. In blue the boundaries of the $z$ domain verifying $z_{\alpha}+z_{\beta} < 1$, $z_{\alpha} < \alpha $ and $z_{\beta} < \beta$. For this figure, $\alpha = 0.8$ and $\beta=0.9$}
	\label{fig:phase}
\end{figure}

The analysis of the single-species TASEP presented earlier can be naturally extended to the multi-species case. The phase diagram is determined by the signs of the characteristic speeds, which correspond to the eigenvalues of the current Jacobian matrix. However, to have a physical interpretation of these phases, it is necessary to use the Riemann variables in the bulk rather than the densities. (For single-component systems, the Riemann variable is simply the density.) Each Riemann variable in the bulk $z_i^B$ can exhibit one of three behaviors:
\begin{itemize}[label=\raisebox{0.2ex}{\tiny$\blacktriangleright$}]
    \item It can be driven from the left boundary $z_i^B = z_i^L$ if the corresponding eigenvalue is positive $v_i(\bfz) > 0$
    \item  It can be driven from the right $z_i^B = z_i^R$ if  $v_i(\bfz) < 0$
    \item It can be bulk-driven (not induced by either boundary) if $v_i(\bfz) = 0$
\end{itemize}
For a system with $N$ species, one expects to have $3^N$ different phases; however, the hyperbolic condition imposes that the characteristic velocities are strictly ordered and thus makes some of the phases forbidden. For the two TASEP, we have 5 phases that can be featured in the $\bfz$ domain. Figure \ref{fig:phase}

\section{Overview of Hyperbolic Systems of Conservation Laws in 1D}
\label{Hyperbolic}
Since solving the Riemann problem is a key component of the method, we provide here a concise review of the fundamental tools required to solve it. This section draws on classical texts such as \cite{evans2010partial,serre1999systems1,serre1999systems2,lax2006hyperbolic,dafermos2005hyperbolic,menonpde}. A particularly useful reference is \cite{bressan2013hyperbolic}.  See also Chapter 1 of \cite{zahra2023multi}.

Consider a set of \(n\) conserved quantities with local densities:
$
\boldsymbol{\rho}(x,t) = (\rho_1(x,t), \dots, \rho_n(x,t)), \quad x \in \mathbb{R}, \quad t \in \mathbb{R}^+,
$
and their associated currents:
$
\boldsymbol{J}(\boldsymbol{\rho}) = (J_1(\boldsymbol{\rho}), \dots, J_n(\boldsymbol{\rho})),
$
where \(J_i\) represents the current associated with \(\rho_i\). These fields evolve according to the coupled partial differential conservation laws:
\begin{equation}
\partial_t \boldsymbol{\rho} + \partial_x \boldsymbol{J} = 0,
\label{eqn:con}
\end{equation}
with the initial condition:
\begin{equation}
\boldsymbol{\rho}(x, 0) = \boldsymbol{\rho}_0(x).
\end{equation}

In general, this system may exhibit discontinuous solutions, and the conservation laws must be interpreted thus in the weak sense. That is, for any smooth test function \(\phi(x,t)\) with compact support and for any \(1 \leq i \leq n\), the following condition must hold:
\begin{equation}
\label{weak}
\int_{\mathbb{R}} \int_{\mathbb{R}^+}
(\rho_i \partial_t \phi + J_i \partial_x \phi) \, dx \, dt = 0.
\end{equation}

The existence and/or uniqueness of solutions for arbitrary initial conditions is a challenging problem and remains an active area of research \cite{bressan2011open}. However, for our purposes, we restrict the analysis to a specific initial condition in the form of a step function at the origin:
$
\boldsymbol{\rho}(x, 0) =
\begin{cases}
\boldsymbol{\rho}^L & \text{if } x < 0, \\
\boldsymbol{\rho}^R & \text{if } x > 0.
\end{cases}
$
The system of conservation laws with this step-function initial condition is known as the \textit{Riemann problem}.
This particular initial condition leads to a key simplification: the problem becomes invariant under the rescaling \((x, t) \to (\lambda x, \lambda t)\), for any $\lambda \neq 0$. Consequently, the solution must also respect this invariance, meaning that \(\boldsymbol{\rho}(x, t)\) can be expressed as a function of a single variable \(\boldsymbol{\rho}(\xi)\), where \(\xi := \frac{x}{t}\). This transformation reduces the PDE to an ODE:
\begin{equation}
\label{cons-xi}
\boldsymbol[-\xi + A(\Brho)] \frac{d \Brho}{ d \xi} = 0,
\end{equation}
where \(A (\Brho)\) is the Jacobian of the current, defined as \(A_{ij} := \frac{\partial J_i}{\partial \rho_j}\).
Let's first briefly discuss the particular case of a single component system.

\subsection{The Scalar Case}
For a single conserved quantity ($n = 1$), the analysis becomes straightforward.First, when the flux is linear, \(J(\rho)=\lambda\rho\), the conservation law reduces to the linear advection equation.  Any initial profile is simply advected at the constant speed \(\lambda\); its shape remains unchanged. For a \emph{nonlinear} flux, one can still reason locally: a small perturbation about a background density \(\rho\) propagates with velocity \(J'(\rho)\), known as the \emph{characteristic speed}.  The corresponding \emph{characteristic curves} are the paths in the \((x,t)\)-plane along which such disturbances travel.

Assume now that $J(\rho)$ is a smooth and convex function, and define $v(\rho) = J'(\rho)$. The governing equation is:
$$
(-\xi + v(\rho)) \rho' = 0.
$$
Beyond the trivial constant solutions, this equation admits solutions of the form:
$$
\rho(\xi) = v^{-1}(\xi).
$$
Thanks to the convexity of $J(\rho)$, the function $v(\rho)$ is invertible, and $v^{-1}$ is strictly increasing. Define $\xi^- := v(\rho^L)$ and $\xi^+ := v(\rho^R)$. If $\xi^- < \xi^+$, a continuous solution can be constructed:

$$
\rho(\xi) = 
\begin{cases}
\rho^L & \text{if } \xi < \xi^-, \\
v^{-1}(\xi) & \text{if } \xi^- < \xi < \xi^+, \\
\rho^R & \text{if } \xi > \xi^+.
\end{cases}
$$

Qualitatively, this means that starting from the Riemann initial condition, the solution evolves as a wave that gradually spreads out from the origin, propagating at a speed $\xi^-$ at the left edge and $\xi^+$ at the right edge. The regions outside the interval $[\xi^- t, \xi^+ t]$ remain unaffected up to time $t$. This type of solution is called a \textit{rarefaction fan}.

In the case where $\xi^- > \xi^+$, a solution of the previous form does not exist, and it becomes necessary to consider the weak formulation of the equation. Conservation laws in this context are known to admit discontinuous solutions called \textit{shock waves}. For the Riemann initial condition, the shock position is given by $x_{\text{sh}}(t) = v_{\text{sh}} t$, where $v_{\text{sh}}$ is the shock speed. At the shock discontinuity, the weak form of the conservation law reduces to the \textit{Rankine-Hugoniot condition}, which must be satisfied:

$$
(\rho^R - \rho^L) v_{\text{sh}} = J(\rho^R) - J(\rho^L).
$$

Here, $\rho^L$ and $\rho^R$ are the densities on the left and right sides of the discontinuity. The solution to the Riemann problem under the condition $\xi^- > \xi^+$ is then:

$$
\rho(\xi) = 
\begin{cases}
\rho^L & \text{if } \xi < v_{\text{sh}}, \\
\rho^R & \text{if } \xi > v_{\text{sh}}.
\end{cases}
$$

\textbf{Remark:} This is a weak solution of the conservation laws, even in the case where $v(\rho^L) < v(\rho^R)$; however, in this case, such a solution would not be stable, meaning that an infinitesimal perturbation can transform the shock into a rarefaction fan or split it into multiple shocks. To ensure stability, the solution must satisfy \textit{admissibility conditions}, which selects physically relevant solutions and guarantees uniqueness. 

One commonly used admissibility criterion is the \textit{Lax condition}, given by:

$$
v(\rho^L) \geq v_{\text{sh}}(\rho^L, \rho^R) \geq v(\rho^R).
$$
imply that these characteristic lines on both sides tilt \emph{into} the shock, so every perturbation is absorbed rather than emitted.
The jump therefore produces entropy—information is irreversibly lost—whereas reversing the inequalities would let characteristics escape, rendering the discontinuity unstable and time-reversible.
\subsection{The Multi-Component Case} \label{cons}
The analysis becomes significantly more intricate when $n$ coupled conserved quantities are present. The Jacobian of the currents, defined as 
$$A(\Brho^{*}) = \frac{\partial \boldsymbol{J}}{\partial \Brho}\big|_{\Brho = \Brho^{*}}$$
plays a crucial role in this scenario. A system of conservation laws is said to be \textit{strictly hyperbolic} if this Jacobian matrix is diagonalizable over $\mathbb{R}$, and its eigenvalues are distinct for all $\Brho$:
\begin{equation}
  \lambda_1(\Brho) < \lambda_2(\Brho) < \dots < \lambda_n(\Brho).  
\end{equation}

Under this condition, that we assume to hold for this review, we can select the left and right eigenvectors ($\mathbf{l}_i$ and $\mathbf{r}_i$, respectively) such that they satisfy the bi-orthogonality condition:
\begin{equation}
    \mathbf{l}_i \cdot \mathbf{r}_j = \delta_{i,j}.
\end{equation}

This condition can also be viewed as equivalent to the decomposition of the Jacobian matrix:
\begin{equation}
    \mathbf{A}(\Brho) = R \mathbf{\Lambda} L,
\end{equation}

where $R$ is the matrix whose columns are the right eigenvectors, $L$ is the matrix whose rows are the left eigenvectors, and $\mathbf{\Lambda}$ is the diagonal matrix of eigenvalues $\lambda_i$. Additionally, the matrices satisfy the relationship $R \cdot L = \mathbf{I}$, where $\mathbf{I}$ is the identity matrix.

\subsubsection{The Linear Case}
To build intuition, it is helpful to start with the case where $A$ does not depend on $\Brho$. In this situation, the system is linear, and the density field $\Brho$ can be expressed as a sum of components aligned with the right eigenvectors:
\begin{equation}
    \Brho = \sum_{i=1}^{n} \tilde{\rho}_i \mathbf{r}_i,
\end{equation}

where the transformed densities $\tilde{\rho}_i$ are given by:
\begin{equation}
    \tilde{\rho}_i = \Brho \cdot \mathbf{l}_i.
\end{equation}
By multiplying both sides of the conservation laws \ref{eqn:con} with $\mathbf{l}_i$, it can be shown that each transformed density satisfies an independent scalar linear conservation law:
\begin{equation}
    (\tilde{\rho}_i)_t + \lambda_i (\tilde{\rho}_i)_x = 0,
\end{equation}

where $\lambda_i$ is the eigenvalue associated with $\mathbf{l}_i$. Consequently, these new quantities evolve independently, and the solution regarding the original density variables at time $t$ is given in terms of the initial density $t=0$ as :
\begin{equation}
   \Brho(x,t) = \sum_{i=1}^{n} (\Brho(x - \lambda_i t, 0) \cdot \mathbf{l}_i) \mathbf{r}_i. 
\end{equation}

\subsubsection{Nonlinear Case}
When $A$ becomes a function of $\Brho$, the analysis grows substantially more complex. The different waves can now interact with each other, leading to nonlinear coupling between the components. This interaction introduces significant challenges in solving the system and understanding the behavior of its solutions. We start by examining the weak solution.

\subsubsection{Weak Solutions and the Rankine-Hugoniot Condition}

As in the scalar case, the conservation equations must be interpreted in their weak form, allowing for discontinuous solutions. A natural extension of the Rankine-Hugoniot condition applies in this context, stating that discontinuities must satisfy:

\begin{equation}\label{RHC}
(\Brho^{+} - \Brho^{-}) \sigma = (\boldsymbol{J}(\Brho^{+}) - \boldsymbol{J}(\Brho^{-})),
\end{equation}
where $\Brho^{+}$ and $\Brho^{-}$ represent the densities on the right and left of the discontinuity, respectively, and $\sigma \in \mathbb{R}$ denotes the shock speed. Unlike the scalar case, this condition not only determines the shock speed but also imposes constraints on the admissible densities between which a shock solution can exist. Specifically, for all $1 \leq i, j \leq n$, the following determinant condition must hold:
\begin{equation}
\text{Det} \begin{pmatrix}
    \rho_{i}^{+} - \rho_{i}^{-} & J_{i}(\Brho^{+}) - J_{i}(\Brho^{-}) \\
    \rho_{j}^{+} - \rho_{j}^{-} & J_{j}(\Brho^{+}) - J_{j}(\Brho^{-})
\end{pmatrix} = 0.
\end{equation}

This constraint introduces the concept of \textit{shock curves}.

\subsubsection{Shock Curves}
Consider a fixed point $\Brho^{*}$ in the density domain, and search for all points $\Brho$ that can be connected to $\Brho^{*}$ via a shock. This set can be parameterized by the shock velocity $\sigma$, forming a one-dimensional manifold in the density domain. In reality, this manifold consists of $n$ distinct  curves passing through $\Brho^{*}$, each is associated with one eigenvalue of the Jacobian matrix.

To see this, we linearize the Rankine-Hugoniot condition near $\Brho^{*}$:

\begin{equation}
\Brho = \Brho^{*} + \frac{1}{\sigma} A(\Brho^{*})(\Brho - \Brho^{*}),
\end{equation}

 Since the jacobian $A(\Brho^{*})$ has $n$ distinct real eigenvalues, this equation admits $n$ independent one-dimensional eigenspaces, each representing the tangent to one of the shock curves. These shock curves can be parameterized as $\sigma \rightarrow S^{i}_{\Brho^{*}}(\sigma)$, where $i$ indicates the curve corresponding to the $i$-th eigenvalue $\lambda_i$.
Note that a shock of an infinitesimal amplitude ( $ \Brho - \Brho^*  \approx 0 $, i.e. a perturbation) along the i-shock curve, has a shock velocity $\sigma = \lambda_i$. This means:
\begin{equation}
S^{i}_{\Brho^{*}}(\lambda_i) = \Brho^{*}
\end{equation}

Note that the $i$-shock curve originating from a given point $\Brho^{*}$ does not, in general, coincide with the $i$-shock curve originating from a different point on the same curve. An exception will be made later for a class of systems called \textit{Temple Class}

\subsubsection{Admissibility Conditions}

Weak solutions are not necessarily unique, so admissibility conditions are needed to select the "physical" solutions. A conceptual approach involves adding a small diffusion term to the conservation equation:

\begin{equation}
\Brho_{t} + \boldsymbol{J}(\Brho)_{x} + \epsilon \Brho_{xx} = 0,
\end{equation}

and then searching for solutions that converge in $L^{1}_{\text{loc}}$ as $\epsilon \to 0$. While this method worked for the Burgers equation, as shown by Hopf \cite{hopf1950partial}, it does not generally offer a practical procedure to eliminate non-physical solutions, necessitating alternative approaches.
\subsubsection{Lax Condition}
Let $\lambda_{i}(\Brho^{L})$ and $\lambda_{i}(\Brho^{R})$ be the $i$-th eigenvalues at the left and right of a discontinuity respectively. The Lax condition is expressed as:
\begin{equation}
\lambda_{i}(\Brho^{L}) \geq \sigma \geq \lambda_{i}(\Brho^{R}),
\end{equation}
indicating that small perturbations along the shock curve on either side of the shock must move toward the shock, ensuring its stability.
This condition embodies the irreversibility of physical solutions: information from the initial data is irretrievably lost in the shock. Mathematical solutions that reverse this process are unphysical.

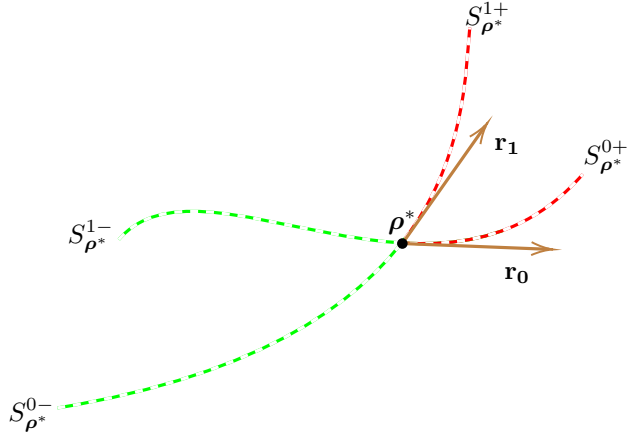
\begin{figure}[h!]
	\centering
	
	\begin{tikzpicture}[x=0.75pt,y=0.75pt,yscale=-1,xscale=1]

	\tikzset{
		bicolor/.style 2 args={
			dashed,dash pattern=on 150pt off 90pt,#1,
			postaction={draw,dashed,dash pattern=on 90pt off 150pt,#2,dash phase=130pt}
		},
	}

	\draw  [very thick,bicolor={green}{red}]  (204,148.5) .. controls (246,99.5) and (365,199.5) .. (438,115.5) ;
	
	\draw [ultra thick,dashed, color = white, very thick]  (204,148.5) .. controls (246,99.5) and (365,199.5) .. (438,115.5) ;
	
	\draw  [very thick, color=brown]  (347,150.5) -- (421,153.42) ;
	
	\draw [very thick,shift={(423,153.5)}, rotate = 182.26] [color=brown ][line width=0.75]
	(10.93,-3.29) .. controls (6.95,-1.4) and (3.31,-0.3) .. (0,0) .. controls (3.31,0.3) and (6.95,1.4) .. (10.93,3.29)   ;
	
	\tikzset{
		bicolor/.style 2 args={
			dashed,dash pattern=on 150pt off 90pt,#1,
			postaction={draw,dashed,dash pattern=on 90pt off 150pt,#2,dash phase=90pt}
		},
	}
	
	\draw [very thick,bicolor={green}{red}]   (173,233.5) .. controls (381,199.5) and (377,84) .. (381,41.5) ;
	
	\draw [ultra thick,dashed, color = white, very thick]  (173,233.5) .. controls (381,199.5) and (377,84) .. (381,41.5) ;
	
	\draw [very thick,color=brown]   (347,150.5) -- (388.85,91.13) ;
	\draw [very thick,shift={(390,89.5)}, rotate = 125.18] [color=brown][line width=0.75]   (10.93,-3.29) .. controls (6.95,-1.4) and (3.31,-0.3) .. (0,0) .. controls (3.31,0.3) and (6.95,1.4) .. (10.93,3.29)   ;
	
	\fill (347,150.5) circle(2pt) node[above]{$ \Brho^{*} $};
	
	\fill (405,175) node[above] {$ \mathbf{r_{0}} $};
	\fill (400,110) node[above] {$ \mathbf{r_{1}} $};
    
	\fill (390,50) node[above] {$ S^{1 +}_{\Brho^{*}} $};
	\fill (450,120) node[above] {$ S^{0 +}_{\Brho^{*}} $};
    
    \fill (190,160) node[above] {$ S^{1 -}_{\Brho^{*}} $};
	\fill (160,250) node[above] {$ S^{0 -}_{\Brho^{*}} $};
    
	\end{tikzpicture}
	\caption{Shock curves originating from $\Brho^{*}$ in a two-component system. Each curve consists of two parts: the admissible side (shown in green) and the non-admissible side (shown in red). The right eigenvectors are tangents to the shock curves and are conventionally oriented toward the non-admissible side.}
	\label{fig:admis}
\end{figure}

Concretely, admissibility conditions eliminate one side of each shock curve originating from $\Brho^{*}$. Denote $S^{i+}_{\Brho^{*}}$ as the non-admissible side and $S^{i-}_{\Brho^{*}}$ as the admissible side.We set the orientation of $\mathbf{r}_{i}$ to be in the direction of the non-admissible side, as shown in Figure \ref{fig:admis}. This orientation will be consistent with further developments.

Recall that in the scalar case, the convexity of the current greatly simplified the analysis, primarily because the mapping from the density to the speed of perturbations is monotonic and, therefore, one-to-one. In higher dimensions, we require a similar property to hold along the integral curves of the eigenvectors of the Jacobian matrix, as we will explore in the following sections.

\subsubsection*{A Simplifying Hypothesis}

A common hypothesis in the study of hyperbolic systems of conservation laws, originating from Lax (1957) \cite{lax1957hyperbolic}, is to assume that each of the eigenvector fields belongs to one of the two categories:
\begin{itemize}
    \item $\nabla_{\Brho} \lambda_{i} \cdot \mathbf{r}_{i}(\Brho) > 0$ for all $\Brho$, indicating that the field is \textit{genuinely nonlinear}.
    \item $\nabla_{\Brho} \lambda_{i} \cdot \mathbf{r}_{i}(\Brho) = 0$ for all $\Brho$, indicating that the field is \textit{linearly degenerate}.
\end{itemize}

For a genuinely nonlinear field, this condition means that the directional derivative of $\lambda_{i}$ along $\mathbf{r}_{i}$ does not change its sign. Whether this sign is positive or negative is a convention that can be decided by the choice of the orientation of $\mathbf{r}_{i}$. We choose this orientation such that the sign is positive. This implies that $\lambda_{i}$ is an increasing function along the directed integral curves of the $i$-field, analogous to the convexity condition in the scalar case. For a linearly degenerate field, $\lambda_{i}$ remains constant along its integral curves.

\subsubsection{Rarefaction Curves}

Consider the integral curves (or field lines) of the eigenvector field $\mathbf{r}_{i}(\Brho)$. For a genuinely nonlinear field, an integral curve can be parameterized by the corresponding eigenvalue field. Thus, the $i$-curve passing through a density point $\Brho^{*}$ can be parameterized as:

\begin{equation}
\lambda_{i} \rightarrow R_{\Brho^{*}}^{i}(\lambda_{i}).
\end{equation}

This curve is called an $i$-\textit{rarefaction curve}. The $i$-rarefaction curve passing through $\Brho^{*}$ is tangent to the $i$-shock curve originating at $\Brho^{*}$ and furthermore shares the same curvature at $\Brho^{*}$ (see Figure \ref{rarefaction}). For a specific class of conservation laws, known as the \textit{Temple class}, these two curves are identical for all fields \cite{temple1983systems}.

The point $\Brho^{*}$ divides the rarefaction curve into two segments. Let $R^{i+}_{\Brho^{*}}$ denote the segment where $\lambda_{i}(\Brho) > \lambda_{i}(\Brho^{*})$, and $R^{i-}_{\Brho^{*}}$ denote the segment where $\lambda_{i}(\Brho) < \lambda_{i}(\Brho^{*})$.

\subsubsection{T-Curves}
\begin{figure}[h!]
	\centering
	
	\begin{tikzpicture}[x=0.75pt,y=0.75pt,yscale=-1,xscale=1]

	\draw [thick, dash pattern=on 5pt off 5pt, draw=black, opacity=0.2, <-] 
	    (204,148.5) .. controls (246,99.5) and (365,199.5) .. (438,115.5);

	\draw [thick, <-] 
	    (204,148.5) .. controls (246,99.5) and (365,199.5) .. (438,115.5);
	\draw [thick, dash pattern=on 120pt off 100pt, dash phase=10pt, draw=brown, opacity=1, <-] 
	    (204,148.5) .. controls (246,99.5) and (365,199.5) .. (438,115.5);

	\draw [ultra thick, dash pattern=on 5pt off 5pt, dash phase=7pt, draw=white, opacity=1, ] 
	    (204,148.5) .. controls (246,99.5) and (365,199.5) .. (438,115.5);

	\draw [very thick] (347,150.5) -- (421,153.42);

	\draw [very thick, shift={(423,153.5)}, rotate = 180]
	[line width=0.75] 
	(10.93,-3.29) .. controls (6.95,-1.4) and (3.31,-0.3) .. (0,0)
	.. controls (3.31,0.3) and (6.95,1.4) .. (10.93,3.29);


    \draw [thick] (175,104.5) .. controls (399,176.5) and (382,161.5) .. (446,72.5);
	\draw [thick, dash pattern=on 120pt off 140pt, dash phase=125pt, draw=brown, opacity=1, ->] (175,104.5) .. controls (399,176.5) and (382,161.5) .. (446,72.5);

	\fill (347,150.5) circle(2pt) node[above]{$ \Brho^* $};
	\fill (405,175) node[above] {$ \mathbf{r_{i}} $};
	\fill (450,120) node[above] {$ S^{i+}_{\Brho^*} $};
	\fill (460,85) node[above] {$ R^{i+}_{\Brho^*} $};
	\fill (250,125) node[above] {$ R^{i-}_{\Brho^*} $};
	\fill (250,160) node[above] {$ S^{i-}_{\Brho^*} $};

	\end{tikzpicture}
	
\caption{An $i$-rarefaction curve (solid line) and an $i$-shock curve (dashed line) originating from the density point $\Brho^*$. The arrows along the curves represent the physical directions. The segments $R^{i+}_{\Brho^{*}}$ and $S^{i-}_{\Brho^{*}}$ can be joined to form a continuous and smooth curve known as a T-curve, coulored in brown.}
	\label{rarefaction}
\end{figure}
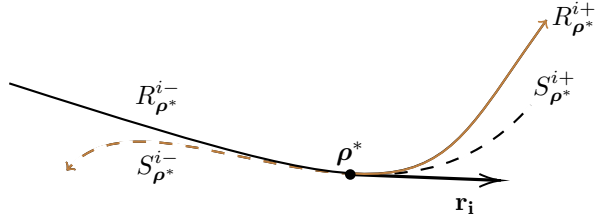

The previous definitions allow us to define a new curve, called a \textit{T-curve}, by stitching together $R^{i+}_{\Brho^{*}}$ with $S^{i-}_{\Brho^{*}}$, Figure \ref{rarefaction}. This curve will play an important role in subsequent analysis.

\subsubsection{The Riemann Problem}
Considering again the equation
\begin{equation}\label{con compact}
(\xi - A) \mathbf{\Brho}'(\xi) = 0,
\end{equation}
with step initial conditions, we can identify the following solutions beyond the trivial constant ones. These are referred to as \textit{elementary solutions}, each consisting of a single type of wave:

\begin{itemize}
    \item \textbf{Shock Solution:}  
    If $\Brho^{L}$ and $\Brho^{R}$ lie on the same $i$-shock curve and satisfy $\lambda_{i}(\Brho^{L}) \geq \lambda_{i}(\Brho^{R})$, the solution to the Riemann problem is a simple shock:

    \begin{equation}
    \Brho(\xi) = 
    \begin{cases}
        \Brho^{L}, & \text{if } \xi < \sigma(\Brho^{L}, \Brho^{R}), \\
        \Brho^{R}, & \text{if } \xi > \sigma(\Brho^{L}, \Brho^{R}).
    \end{cases}
    \end{equation}

    \item \textbf{Rarefaction Wave:}  
    If $\Brho^{L}$ and $\Brho^{R}$ lie on the same genuinely nonlinear $i$-rarefaction curve and satisfy $\lambda_{i}(\Brho^{L}) < \lambda_{i}(\Brho^{R})$, the solution to the Riemann problem is a simple rarefaction wave:

    \begin{equation}
    \Brho(\xi) = 
    \begin{cases}
        \Brho^{L}, & \text{if } \xi < \lambda_{i}(\Brho^{L}), \\
        R_{\Brho^{L}}^{i}(\xi), & \text{if } \lambda_{i}(\Brho^{L}) < \xi < \lambda_{i}(\Brho^{R}), \\ 
        \Brho^{R}, & \text{if } \xi > \lambda_{i}(\Brho^{R}).
    \end{cases}
    \end{equation}

    To confirm why the middle branch holds, multiply Equation~\eqref{con compact} on the right by $\mathbf{r}_{i}$. This leads to an eigenvalue equation, which is solved by the rarefaction branch.

    \item \textbf{Contact Discontinuity:}  
    If $\Brho^{L}$ and $\Brho^{R}$ lie on the same linearly degenerate $i$-rarefaction curve where the eigenvalue $\lambda_{i}$ is constant, the solution to the Riemann problem is a shock:

    \begin{equation}
    \Brho(\xi) = 
    \begin{cases}
        \Brho^{L}, & \text{if } \xi < \lambda_{i}, \\
        \Brho^{R}, & \text{if } \xi > \lambda_{i}.
    \end{cases}
    \end{equation}

    This type of shock is often referred to as a \textit{contact discontinuity}. Unlike the previous cases, a linearly degenerate curve is bidirectional, simultaneously serving as both a rarefaction and a shock curve.
\end{itemize}

In conclusion, for a physical elementary solution, $\Brho^{R}$ must lie on one of the T-curves originating from $\Brho^{L}$, or vice versa.

\subsubsection{Combined Solutions}

If neither $\Brho^{R}$ nor $\Brho^{L}$ is located on the T-curves originating from the other, it is possible to have a solution connecting both by combining elementary solutions from different T-curves. In general, for a system with $n$ components, this requires a combination of $n$ T-curves. Such a combination always exists and is unique, provided that $\Brho^{L}$ and $\Brho^{R}$ are sufficiently close. For the two-TASEP, these combined solutions are represented here in the Riemann variables plane, figure \ref{fig:sol}. A representation in the density plane can be found in \cite{cantini2022hydrodynamic}.

\subsection{Riemann Variables}
Consider the vector field of the left eigenvectors, satisfying the eigenvalue equation:
\begin{equation}
\mathbf{l}_{i} A = \lambda_{i} \mathbf{l}_{i}.
\end{equation}
Assume that, up to a scalar multiplicative factor, $\mathbf{l}_{i}$ can be derived from a scalar potential $z_{i}(\Brho)$, i.e., $\mathbf{l}_{i} \propto \nabla z_{i}$. We can normalize $\mathbf{l}_{i}$ such that:

\begin{equation}
\mathbf{l}_{i} = \nabla z_{i}.
\end{equation}

These scalar fields $z_{i}$ are called the \textit{Riemann variables}. Riemann variables always exist for $n = 2$, but for $n > 2$, they do not generally exist.

Riemann variables simplify the analysis of conservation laws by partially decoupling the system. Specifically, we have:

\begin{equation}
\begin{split}
\partial_{t} z_{i}(\Brho) + \lambda_{i} \partial_{x} z_{i}(\Brho) &=
\nabla z_{i} \cdot \partial_{t} \Brho + \lambda_{i} \nabla z_{i} \cdot \partial_{x} \Brho \\
&= \nabla z_{i} \cdot (\partial_{t} \Brho + \lambda_{i} \partial_{x} \Brho) \\
&= \mathbf{l}_{i} \cdot (\partial_{t} \Brho + \lambda_{i} \partial_{x} \Brho) \\
&= \mathbf{l}_{i} \cdot (\partial_{t} \Brho + A \partial_{x} \Brho) = 0.
\end{split}
\end{equation}

This implies that when the system is expressed in terms of Riemann variables, the coupling between equations is restricted to the velocity coefficients $\lambda_{i}(\mathbf{z})$.

The surfaces where $z_{i}$ is constant form a foliation of $(n-1)$-dimensional manifolds in the density space. These surfaces are perpendicular to the integral curves of $\mathbf{l}_{i}$. Since $\mathbf{l}_{i} \cdot \mathbf{r}_{j} = \delta_{i,j}$, it follows that $z_{i}$ is constant along all $j$-rarefaction curves for $j \neq i$. For this reason, $z_{i}$ are sometimes referred to as the \textit{Riemann invariants}.

The characteristic speeds $\lambda_{i}(\boldsymbol{z})$ can be computed using:

\begin{equation}
\lambda_{i}(\boldsymbol{z}) = \frac{\frac{\partial J_{k}}{\partial z_{i}}}{\frac{\partial \rho_{k}}{\partial z_{i}}}.
\end{equation}

This holds for any choice of $k$ and implies the consistency condition:

\begin{equation}
\frac{\partial J_{n}}{\partial z_{i}} \frac{\partial \rho_{m}}{\partial z_{i}} = \frac{\partial J_{m}}{\partial z_{i}} \frac{\partial \rho_{n}}{\partial z_{i}}.
\end{equation}
Using the explicit expressions collected in Appendix~\ref{Annex} one can
check that this identity indeed holds for the two-TASEP.

\medskip
Although the existence of Riemann variables greatly simplifies a system of
conservation laws—each component obeys the scalar-like equation
$$
\partial_{t} z_{i}\;+\;\lambda_{i}(\boldsymbol{z})\,
\partial_{x} z_{i}=0,
$$
the system is \emph{not} fully decoupled: every characteristic speed
$\lambda_{i}$ still depends on \emph{all} Riemann variables, not solely on
$z_{i}$.

\section{Conclusion}\label{sec:conclusion}

Hydrodynamic methods have long been applied to driven diffusive systems with
open boundaries, yet for multi-species models they were mostly limited to cases
in which the boundary densities are \emph{a priori} known functions of the
boundary rates.  
We have presented a more general framework.  Its key idea is to couple  
(i) the stationarity conditions for the stationary currents with  
(ii) a consistency relation that ties the boundary densities to the bulk
density derived by solving the Riemann problem at the origin.  
Together these relations form a closed system that determines bulk and
boundary densities simultaneously.

The method neither requires integrable boundaries nor assumes a product-form local
invariant measure; the only prerequisite is the knowledge of the hydrodynamic
currents for the spatially uniform system (e.g.\ on a ring).  

Applied to the two-species TASEP with arbitrary bulk hopping rates and
non-integrable boundaries, the scheme yields  
\begin{enumerate}[label=(\roman*),leftmargin=*]
  \item bulk and boundary densities for any choice of rates, and  
  \item a five-phase diagram whose regions are distinguished solely by the
        signs of the characteristic velocities.  
\end{enumerate}
The same strategy was shown in~\cite{cantini2024steady} to work for other
two and three-species models. 
All examples studied so far admit Riemann variables, which greatly simplify the
analysis.  A natural next step is to treat models with three or more species
\emph{without} such variables.  We expect that bulk and boundary densities can
still be determined, but whether the corresponding phase diagram retains an
equally transparent interpretation remains an open question.  
It would also be interesting to apply this method to a broader class of models beyond exclusion processes, such as the multi-species zero-range process studied in~\cite{grosskinsky2003stationary}.  
Finally, preliminary evidence suggests that Temple-class models may display unusually
simple boundary-driven behaviour; investigating this conjecture is another
promising direction for future work.

\appendix

\section{Explicit formulas for the Two-TASEP}
\label{Annex}

The Riemann variables as functions of the densities
\begin{align}
z_{\alpha}
=
\frac{1}{2} \left(1 - \rho_{\bullet} + \alpha (\rho_{\circ} + \rho_{\bullet}) - \sqrt{-4 \alpha \rho_{\circ} + \left(1 - \rho_{\bullet} + \alpha (\rho_{\circ} + \rho_{\bullet})\right)^2}\right) \\
z_{\beta}
=
\frac{1}{2}
\left(1 - \rho_{\circ} + \beta (\rho_{\circ} +\rho_{\bullet}) - \sqrt{-4 \beta \rho_{\bullet} + \left(1 - \rho_{\circ} + \beta (\rho_{\circ} +\rho_{\bullet})\right)^2}\right)
\end{align}
The previous relations are invertible in the interior of the physical domain 
\begin{align}
\rho_{\circ}(z_{\alpha},z_{\beta}) &=
\frac{z_{\alpha} (1-z_{\beta}) (\beta (1 - z_{\alpha}) - z_{\beta}(1-\alpha))}
{z_{\alpha} z_{\beta} (\alpha + \beta - 1) - \alpha \beta (1 - z_{\alpha} - z_{\beta})}
\\
\rho_{\bullet}(z_{\alpha},z_{\beta}) &=
\frac{z_{\beta}(1 - z_{\alpha}) ( \alpha (1 - z_{\beta}) - z_{\alpha}(1-\beta) )}
{z_{\alpha} z_{\beta} (\alpha + \beta - 1) - \alpha \beta (1 - z_{\alpha} - z_{\beta})}
\end{align}

The currents as functions of the Rieman variables
\begin{align}
J_0
=
\frac{z_{\alpha} (z_{\beta} - 1) \left[ \beta (z_{\alpha} - 1) (z_{\alpha} - \alpha) - (\alpha - 1) \beta z_{\beta} + (\alpha - 1) z_{\beta}^2 \right]}{z_{\alpha} z_{\beta} (\alpha + \beta - 1) - \alpha \beta (1 - z_{\alpha} - z_{\beta})}
 \\
J_1
=
-\frac{(z_{\alpha} - 1) z_{\beta}
\left[\alpha (z_\beta-1)(z_\beta - \beta ) - (\beta-1) \alpha z_\alpha   +(\beta-1) z_\alpha^2\right]}
{z_{\alpha} z_{\beta} (\alpha + \beta - 1) - \alpha \beta (1 - z_{\alpha} - z_{\beta})}
\end{align}

We can notice that the equations are invariant under the transformation:
$$(\alpha,\beta,z_\alpha,z_\beta,J_0,J_1) \rightarrow (\beta,\alpha,z_\beta,z_\alpha,-J_1,J_0)$$
This symmetry originates from the microscopic model.

\begin{equation}
    v_{\alpha} = z_{\alpha} -z_{\beta} + \frac{\rho_{\circ} + z_{\beta}-1}{\partial_{z_{\alpha}} \rho_{\circ}}
\end{equation}

\begin{equation}
    v_{\beta} = z_{\alpha} -z_{\beta} + \frac{ 1 -z_{\alpha}-\rho_{\bullet}}{\partial_{z_{\beta}} \rho_{\bullet}}
\end{equation}

With
\begin{equation}
\frac{\partial \rho_{\circ}}{\partial_{z_{\alpha}}}
   = \frac{\beta (1 - z_\beta) \left[ -\alpha \beta (1 - z_\alpha)^2 + (\alpha (1 - \alpha + \beta) - 2 \alpha \beta z_\alpha + (1 - \alpha - \beta) z_\alpha^2) z_\beta + (1 - \alpha) \alpha z_\beta^2 \right]}{\left[\alpha \beta (1 - z_\alpha) + (\alpha \beta + z_\alpha - (\alpha + \beta) z_\alpha) z_\beta \right]^2}
\end{equation}

\begin{equation}
\frac{\partial \rho_{\bullet}}{\partial_{z_{\beta}}}
   =
    \frac{(1 - \beta) z_\beta \left[ \alpha \beta (1 - z_\alpha)^2 + (\alpha (1 - \alpha - \beta) + 2 \alpha \beta z_\alpha - (1 - \alpha - \beta) z_\alpha^2) z_\beta - (1 - \alpha) \alpha z_\beta^2 \right]}{\left[ \alpha \beta (1 - z_\alpha) + (\alpha \beta + z_\alpha - (\alpha + \beta) z_\alpha) z_\beta \right]^2}
\end{equation}

\begin{equation}
    v_{\beta} = z_{\alpha} -z_{\beta} + \frac{z_{\alpha}-\rho_{\circ}}{\partial_{z_{\beta}} \rho_{\circ}}
\end{equation}
or
\begin{equation}
    v_{\beta} = z_{\alpha} -z_{\beta} + \frac{ 1 -z_{\alpha}-\rho_{\bullet}}{\partial_{z_{\beta}} \rho_{\bullet}}
\end{equation}

Additional formula

\begin{equation}
     \frac{ \partial_{z_{\beta}} \rho_{\circ} }{\partial_{z_{\beta}} \rho_{\bullet}} = \frac{z_{\alpha}-\rho_{\circ}}{1 -z_{\alpha}-\rho_{\bullet}}
\end{equation}

\subsection*{Acknowledgements}
This text is based on a talk given at Institut Henri Poincaré in Paris on the
occasion of the conference Inhomogeneous Random Systems (IRS) 2024. I wish to express
my gratitude to the organisers: Ellen Saada, 
Giambattista Giacomin and Christian Maes for the invitation as well as to Gunter Schutz for organizing a day devoted to Boundaries in driven systems. This work was supported by FCT (Portugal) 
through project UIDB/04459/2020, 
doi 10-54499/UIDP/04459/2020,
by the FCT Grants 2020.03953.CEECIND and 
2022.09232.PTDC, and by ANR-PRME Uniopen, project ANR-22-CE30-0004-01.

\typeout{}
\bibliography{f}
\bibliographystyle{ieeetr}

\end{document}